\documentclass[11pt,a4paper]{article}
\usepackage[margin=1in]{geometry} 
\usepackage{graphicx} 
\graphicspath{{K:/4114/Project/}}
\usepackage{caption}
\usepackage{listings}
\usepackage{xparse}
\usepackage{setspace}
\usepackage{amsmath}
\usepackage{blindtext}
\usepackage{apacite}
\usepackage{natbib}
\usepackage[graphicx]{realboxes}
\usepackage[toc,page]{appendix}
\usepackage{float}
\usepackage{threeparttable}
\usepackage{enumitem}
\usepackage{amsmath,amsfonts,amssymb,amsthm}
\usepackage{url}
\usepackage{longtable,tabu}
\usepackage[super]{nth}

\usepackage{chngcntr}
\usepackage{multirow}

\usepackage{titlesec}
\titleformat{\chapter}[display]
  {\headingfont\centering}{\thechapter}{0pt}{\Huge\textbf}

\usepackage{amsmath}

\title{\LARGE Estimating The Effect Of Subscription based Streaming Services On The Demand For Game Consoles\footnote{ECON 4114 Final Project, HKUST Fall Semester 2018-19}}
\date{Dec 30 2018}
\author{\normalsize Chan Tung Yu Marco  \and \normalsize Zhang Yue  \and \normalsize Yeung Tsun Yi}

\spacing{1.25}

\begin{document}

\maketitle

\begin{abstract}
\normalsize

In this paper, we attempt to estimate the effect of the implementation of subscription based streaming services on the demand of the associated game consoles. We do this by applying the BLP demand estimation model proposed by  \cite{berry1994estimating}. This results in a linear demand specification which can be identified using conventional identification methods such as instrumental variables estimation and fixed effects models. We find that given our dataset, the two-stage least squares (2SLS) regression provides us with convincing estimates that subscription-based streaming services does have a positive effect on the demand of game consoles as proposed by the general principle of complementary goods.   

\end{abstract}

\section{Introduction}
Subscription based business models have been around since the \nth{17} Century which were used by publishers of books and periodicals \citep{clapp1931beginnings}. We also see this in newspapers and magazines, and more recently we are starting to see this being incorporated into digital media platforms; a prime example being the American based media-services provider of TV shows and movies Netflix. \cite{glazer1982economics} shows that subscription services together with individual unit sales allow monopolists to implement price discrimination resulting in a increased profits and net welfare gain. \

However, our topic of interest is investigating how a subscription based service will affect the demand of a complementary good. In the years surrounding 2013, the leaders of the gaming industry Sony, Microsoft and Nintendo released several game consoles, few of which are still leading the game console markets to this day, the notable ones being XBox One (XBO) and PlayStation 4 (PS4).\footnote{More precisely their hardware revised variants} In 2014, Sony released a cloud gaming subscription service PlayStation Now (PS Now) for PlayStation 4 and PC players to pay for access to a selection of PlayStation 2, PlayStation 3 and PlayStation 4 games through a monthly subscription\footnote{https://www.playstation.com/en-us/explore/playstation-now/}. Similarly and more recently in 2017, Microsoft introduced the Xbox GamePass, a similar game-streaming service allowing Xbox One players to gain access to a catalog of games for a monthly subscription fee\footnote{https://www.xbox.com/en-us/xbox-game-pass/}. \

The goal of this study is to investigate the effects of the introduction of such subscription-based streaming services on the demand of the associated game consoles. Using basic economic reasoning on the behaviour of complementary products, our prior would be that since these subscription streaming services are intended to offer a discount on games, then the increased demand due to the discount would have a spillover effect onto the demand on consoles associated with such streaming services. We will be investigating this empirically by estimating the demand for each console where the presence of a subscription based service will be represented by a dummy variable $Subscribe$. Thus, the coefficient on $Subscribe$ will ideally capture its effect on the demand of the respective console.

\section{Econometric Model}

\subsection{Empirical Strategy}

We will be estimating demand using the BLP Discrete-Choice Demand Model  \citep{berry1994estimating}. We assume that the choice probabilities take a multinomial logit form, thus we have to incorporate the \textbf{Independence of Irrelevant Alternatives (IIA)} assumption\footnote{The ratio of the probability of choosing either of of the two alternatives are independent of the set of alternatives available.}. The objective is to obtain structural estimates for the demand function as a function of console characteristics, and hence obtain the coefficient on the dummy variable $Subscribe_{jt}$\footnote{Where 
$\begin{cases}
Subscribe_{jt} = 1, \text{If Subscription service is introduced for console $j$ at time $t$.} \\
Subscribe_{jt} = 0, \text{otherwise}
\end{cases}
$} in order to capture the effect on demand after introducing a subscription-based streaming service.  

\subsection{Data}
The data is in a panel data format with each console $j$ and their product characteristics $x_{jt}$ at a given year $t$. There were a few simplifications that need to be made when constructing our dataset due to a number of complications.\footnote{For example, the market share data for each console does not distinguish between different models of the same console resulting from a number of hardware revisions. To have the regression run more smoothly, each year we assumed that the newer revised model replaces the older model and thus represents the aggregate market share for the year.} Table \ref{table: Definitions} Below shows the definitions of the product characteristics used in the demand estimation.
~\\
\indent The data is obtained from various sources, mainly from different sites that track prices of video game products for video game enthusiasts. The source of market share data is from the site \cite{vg1} accessed through the online statistics portal Statista. Sources of data on product characteristics include product comparison articles on Forbes\footnote{\cite{forbes}} and GameSpot\footnote{\cite{gamespot}}. 
~\\
\indent Data for the costs of components of consoles are obtained from online articles published by IHS Markit, a London-based global information provider. The data comes in the form of bills of materials (BOM) available from press releases. For others where such data is unavailable, we had to use some proxy estimates using substitute components as the price of the specific contraption is not available.\footnote{This is the case for a number of consoles (PS4, NN3DS XL, PS Vita, NS). We had to find substitute CPU processors from the same manufacturer of similar specifications. E.g. the PS4 uses the 2.1 GHz 8-core AMD custom Jaguar. We use the price of the 3.3 GHz 8-core AMD FX-8300 CPU instead which is suggested as a perfect substitute.}

\begin{table}[ht!] 

\centering

\caption{Definitions}
\label{Definitions}
\begin{threeparttable}
\begin{tabular}{c c c}

\hline\hline
Trait & & Definition \\ 
\hline

Vol & & Volume of Game console in $mm^3$ \\
Alone & & Dummy variable for Standalone consoles \\
grams & & Weight in grams (g) \\
Storage & & Storage space in Gigabytes (GB) \tnote{3} \ \tnote{11} \\
Titles & & Number of game titles for respective console in respective year \tnote{2}\\
Exclusive & & Number of exclusive game titles for respective console in respective year \tnote{2}\\
RAM & & RAM capacity of game console in Megabytes (MB)\tnote{3} \ \tnote{11}\\
CPU & & Clock rate of game console in Megahertz (Mhz) \tnote{3} \ \tnote{11} \\
GPU & & GPU processor clocking speed in Megahertz (Mhz) \tnote{3} \ \tnote{11} \\
Core & & Number of processor cores in the CPU \tnote{3} \ \tnote{11} \\
Subscribe & & Dummy variable for presence of Subscription-based Game streaming services \\
log & & Log differences in market share of $j$ and outside alternative \tnote{9} \ \tnote{10}\\
Price & & Console price \tnote{1} \\
CPU\_cost & & CPU component cost \tnote{4} \ \tnote{5} \ \tnote{6} \ \tnote{7} \ \tnote{8} \\
RAM\_cost & & Memory component cost \tnote{4} \ \tnote{5} \ \tnote{6} \ \tnote{7} \ \tnote{8} \\

\hline
\end{tabular}

\begin{tablenotes}

\item[1] \scriptsize{Instead of using the official prices released by each respective company, we use the prices derived from PriceChartering, a website that tracks prices of video game goods sold on eBay and their Marketplace using their proprietary algorithms. We use these prices as they are more varied and create for more variation in the dataset to allow for a smoother regression. All prices for each year are taken from the month November; \citet{price2}}

\item[2] \scriptsize{Aggregate data is retrieved from Wikipedia, with various individual sources.}

\item[3] \scriptsize{\cite{forbes}, Forbes}

\item[4] \scriptsize{\cite{ihs1}, IHS}

\item[5] \scriptsize{\url{https://www.alibaba.com/product-detail/512MB-Quad-Core-wifi-Bluetooth-Banana_60794203127.html?spm=a2700.7724838.2017115.43.69e63a8bnKabTb}
}; proxy estimate for New 3DS XL CPU.

\item[6] \scriptsize{\cite{gamernexus}}

\item[7] \scriptsize{\url{https://www.mouser.com/Semiconductors/Integrated-Circuits-ICs/Embedded-Processors-Controllers/Microprocessors-MPU/ARM-Cortex-A9-Core/_/N-a86qz?P=1yzay25Z1yor86gZ1yzolqg}; proxy estimate for PS Vita and PS Vita Slim CPU}

\item[8] \scriptsize{\url{https://www.mouser.com/ProductDetail/NXP-Freescale/LS1043AXE7MQB?qs=sGAEpiMZZMsk5yEqv3Bk8YnbB1mhTs5HcinEL} \\ \url{SxtuxdUM4foXfRtoQ%3d%3d}}; proxy estimate for Nintendo Switch CPU.

\item[9] \scriptsize{\cite{statistaMS}}

\item[10] \scriptsize{\cite{vg1}}

\item[11] \scriptsize{\cite{trust}}

\end{tablenotes}
\end{threeparttable}
\label{table: Definitions}
\end{table}

\subsection{BLP Demand Estimation Setup}
\begin{itemize}
	\item We have $T=5$ time periods starting from 2014 to 2018.
	\item $N_t$ individual consumers at time $t$ such that $\sum_{j=1}^{J} q_{jt}=N_t$. 
	\item We have $J=5$ alternatives (PS4, Xbox One, Nintendo Switch, Nintendo 3DS, PlayStation Vita).
	\item We have a $K$ vector of characteristics for all consoles $j$ at time $t$, $x_{jt}=(x_{jt1},...,x_{jtK})$. Note that $Subscribe_{jt}$ is an entry in this vector.
	\item We have prices of console $j$ at time $t$, $p_{jt}$ 
	\item We have empirical (observed) market shares $\hat{s}_{jt} = \frac{q_{jt}}{N_t}$. We denote outside market share to be $\hat{s}_{0t}$ to account for the difference (this should be the demand for gaming consoles other than the three alternatives in the same time period). 
	\item Unobserved market/product heterogeneity\footnote{E.g., unobserved characteristics of console, measurement error in price, demand shocks etc.} $j$ at time $t$, $\xi_{jt} = \xi_j + \xi_t + \Delta\xi_{jt}$, where $\xi_j$ is console specific shock, $\xi_t$ is common shock at time $t$, $\Delta\xi_{jt}$ is non-fixed product/product unobserved shocks, specific to $j$ and $t$. 
\end{itemize}

\subsection{Estimation}\label{Estimation}

We first write our indirect random utility for individual $i$ as:

\begin{equation}
\label{indirect random utility}
u_{ijt} = x'_{jt}\beta-\alpha p_{jt} + \xi_{jt} + \varepsilon_{ijt} \equiv \delta_{jt} + \varepsilon_{ijt}  
\end{equation} 

where $\delta_{jt}=x'_{jt} \beta-\alpha p_{jt}+\xi_{jt}$ is the mean utility for console $j$ at time $t$. By writing the indirect random utility in the form of (\ref{indirect random utility}), we are making two important assumptions. 

\begin{enumerate}
	\item We assume $\varepsilon_{ijt}$ to be i.i.d type I extreme value distributed under the logit assumption.
	\item We assume that all consumer heterogeneity will be captured by the term $\varepsilon_{ijt}$, resulting in the homogenous parameters $(\alpha,\beta)$ which we are to estimate.\footnote{This is opposed to the Random Coefficients Model where we estimate $(\alpha_i, \beta_i)$. We assume homogenous parameters to simplify our analysis.}
\end{enumerate}

Under these two assumptions we are able to obtain the closed form for estimated market share $\tilde{s}_{jt}$ as a function of $(\delta_{0t},...,\delta_{Jt})$. (See Appendix \ref{Appendix A}).

\begin{equation}
\tilde{s}_{jt}(\delta_{0t},...,\delta_{Jt}) = \frac{\exp(\delta_{jt})}{\exp(\delta_{1t})+...+\exp(\delta_{Jt})}
\label{estimated market share closed form}
\end{equation}

Then for our estimation, we will be using the following moment conditions:

\begin{equation}
  \begin{cases}
               \hat{s}_{1t} = \tilde{s}_{1t}(\delta_{0t},...,\delta_{Jt}) \\
               \vdots \\
               \hat{s}_{Jt} = \tilde{s}_{Jt}(\delta_{0t},...,\delta_{Jt}) 
               
            \end{cases}
 			\label{market share moment conditions}
 			\
\end{equation}

where we normalize $\delta_{0t}=0$. Then from the moment conditions (\ref{market share moment conditions}) we can see that $\hat{s}_{jt}$ is log-linear and we can write, 

\begin{equation}
  \begin{cases}
               \delta_{0t} = 0 \\
               \delta_{1t} = \log \hat{s}_{1t} - \log \hat{s}_{0t} \\
               \vdots \\
               \delta_{Jt} = \log \hat{s}_{Jt} - \log \hat{s}_{0t} \\ 
               
            \end{cases}
 			\label{log-linear form}
 			\
\end{equation}

where $\log \hat{s}_{0t} = -\log(\exp \delta_{1t}  +...+\exp \delta_{Jt} )$.\footnote{$\log \hat{s}_{0t}$ is interpreted as the log of the `outside market share', so the market share of products not included in the analysis.} Now we have obtained a linear reduced form that can be used for estimation, so we can run OLS on:

\begin{equation}
\log \hat{s}_{jt} - \log \hat{s}_{0t} = x_{jt}\beta + \alpha p_{jt} + \xi_{jt}
\label{Berry's obvious OLS}
\end{equation} 

However, notice that by the definition of the error term $\xi_{jt}$, it is reasonable to assume that $cov(p_{jt},\xi_{jt})\neq 0$. Hence to obtain consistent estimates for $(\alpha,\beta)$ we can either carry out IV estimation that requires valid instruments, or run a fixed effects model.

\subsection{Empirical Estimates and Results}
\subsubsection{Fixed Effects Estimates}
We have the linear specification (\ref{Berry's obvious OLS}) to estimate:

$$
\log \hat{s}_{jt} - \log \hat{s}_{0t} = x_{jt}\beta + \alpha p_{jt} + \xi_{jt}
$$

One method that is available to us for identification since we have a panel data format is to run a two-way fixed effects model on (\ref{Berry's obvious OLS}) to account for time and individual fixed effects. However this still requires that $cov(p_{jt},\Delta \xi_{jt}) = 0$ in order to obtain consistent estimates. It is also implicitly required that although $T$ may be required to be small, $J$ is required to be large. \\
\indent Since our panel data only consists of $J=5$ alternatives, the estimates are shown to have very high standard errors resulting in low levels of significance. Furthermore, as shown in specification (4) in Table \ref{FE two ways}, the coefficient on $Subscribe$ changes drastically compared to the rest of the specifications showing a negative sign. Due to the small magnitude and insignificance of most of the coefficients, these estimates are unlikely to be interpretable and no statistical inference or insight can be made. It should also be noted that some of the specifications were not permitted as the characteristic $Subscribe$ would be dropped due to high levels of multicollinearity with other product characteristics.

\subsubsection{Two-Stage Least Squares Estimates (2SLS)}
Another way to obtain consistent estimates for (\ref{Berry's obvious OLS}) is to run a 2SLS using valid instruments. Such instruments $z_1,...,z_m$ are valid if they satisfy the following assumptions: 

\newtheorem{condition}{Condition}

\begin{condition}[Instrument Relevance] 
\label{parallel trend}
The instrument has a sufficiently high correlation with the endogenous regressor.
\\ \indent i.e.
\[
cov(z_{jt},p_{jt})\neq 0
\label{instrument relevence}
\tag{2}
\]

\end{condition}

\begin{condition}[Instrument Exogeneity] 
\label{parallel trend}
The instrument is uncorrelated with the error term $\xi_{jt}$. 
\\ \indent i.e.
\[
cov(z_{jt},\xi_{jt})= 0
\label{instrument exogeneity}
\tag{2}
\]

\end{condition}

\cite{berry1994estimating} suggests using cost shifters (i.e. Supply shifters), as they should be independent of unobservable market/product characteristics that may affect a consumer's utility. \\
\indent The instruments that we will be using are the cost of the CPU and RAM components of the consoles. Because these are cost factors, it should be reasonable to claim that $cov(\texttt{CPU\_cost},\texttt{CPU}) \neq 0$, $cov(\texttt{CPU\_cost},\texttt{Core}) \neq 0$ and $cov(\texttt{RAM\_cost},\texttt{RAM}) \neq 0$ for the relevance condition. We empirically test these conditions in the next section. 
\\ \indent Looking at the estimates in Table \ref{2SLS}, all specifications show to have highly significant estimates for $Subscribe$\footnote{All estimates are significant at the 99$\%$ level.} as well as being similar in value and having an intuitive positive sign. Out of the four specifications, specification (1) is the only one with a significant coefficient on $Price$ at the 95$\%$ confidence level as well as having an intuitive negative sign. Furthermore, all coefficients are shown to be highly significant. Overall, the estimates from Table \ref{2SLS} make a convincing case that the estimates on the coefficient for $Subscribe$ may be robust; as well as specification (1) being the best estimate for the linear demand form (\ref{Berry's obvious OLS}).

\subsubsection{Testing Instrument Relevance}
\indent To further confirm the validity of the estimates provided by specification (1) we proceed to test the two conditions of instrument validity. The relevance condition can be empirically tested by the conditional $F$ test on the null $H_0:\pi_1=\pi_2=0$.\footnote{Where $\pi_1,\pi_2$ are the first stage coefficients of instruments \texttt{CPU\_cost} and \texttt{RAM\_cost} respectively} Tables \ref{F test (1)} to \ref{F test (4)} show the conditional $F$ tests of all four specification from Table \ref{2SLS}. We see that none of the specifications are able to satisfy the rule of thumb requirement of $F \geq 10$\footnote{This may be an indication of weak instruments which may result in severely biased estimates depending on the severity.} \citep{stock2003introduction}, although the specifications closest to satisfying this requirement are specification (1) that yields $F=8.712$ and (4) that yields $F=9.222$.  

\subsubsection{Testing Instrument Exogeneity}
\indent We continue to test the instrument exogeneity condition on specification (1). We do this by running Sargan's $J$ test \citep{sargan1958estimation}. To run Sargan's $J$ test\footnote{Note that Sargan's $J$ test can only be implemented if the 2SLS model is over-identified, that is $m>k$ for $m$ instruments and $k$ endogenous regressors.}, we run the following regression using the 2SLS residuals:

\begin{equation}
\label{residual regression}
\hat{u}^{2SLS} = \beta_0 + \beta_1 \texttt{CPU\_cost} + \beta_2 \texttt{RAM\_cost} + \beta_3 x_1 + ... + \beta_K x_K + e
\end{equation}

Then obtaining the $F$ statistic from Table \ref{residual regression}, the $J$ statistic is defined to be $J=mF$, therefore $J=0.63$. Because $J \sim \chi^2(m-k)$, we have that the 95$\%$ critical value for $\chi^2(1)$ is $0.0039$, therefore allowing us to accept the null.\footnote{If the null were rejected, that would indicate that there is at least one endogenous instrument.}

\section{Discussion and Conclusion}
The data used in this study involved $J=5$ alternatives for game consoles across $T=5$ years of observations, totalling $22$ observations.\footnote{Nintendo Switch only started to appear in 2017} After obtaining the linear specification as specified from the BLP demand estimation model, we employed both two-way Fixed effects (FE) model and two-stage least squares (2SLS) using Instrumental Variables to attempt identify the effect of $Subscription$ on the demand for consoles. The estimates from the FE model that attempted to control for unobserved individual and time heterogeneity were uninterpretable most likely due to the small number of observations of the panel data. \\
\indent The 2SLS estimates on the other hand yielded significant estimates which were convincingly robust as they varied little across specifications as well as having intuitive signs. Although the Sargan $J$ test supports the exogeneity of specification (1) from Table \ref{2SLS}, the $F$ statistic for testing instrument relevance did not meet the $F\geq10$ cut-off although it is still reasonably close. If we were to interpret the positive coefficient of $1.268$ on $Subscription$ from Table \ref{2SLS}, this represents the positive change as a result of implementing a subscription based streaming service on the demand of game consoles in terms of the log differences of market shares as defined by the BLP demand model. 
\\
\indent To further improve this study, increasing the number of alternatives in the model as well as increasing the number of years (or frequency) of observation may allow for a fixed effects model to run smoothly where estimates can then be compared with IV estimates. Implementing the random coefficient model as mentioned in Section \ref{Estimation} may also yield more structurally robust estimates as the homogenous parameter assumption is relaxed.

\clearpage
\section{Tables}

\begin{table}[!htbp] \centering 
  \caption{Fixed Effects model, Two ways} 
  \label{FE two ways} 
\begin{tabular}{@{\extracolsep{5pt}}lcccc} 
\\[-1.8ex]\hline 
\hline \\[-1.8ex] 
 & \multicolumn{4}{c}{\textit{Dependent variable:}} \\ 
\cline{2-5} 
\\[-1.8ex] & \multicolumn{4}{c}{log} \\ 
\\[-1.8ex] & (1) & (2) & (3) & (4)\\ 
\hline \\[-1.8ex] 
 Vol & $-$0.000 &  &  & 0.00000 \\ 
  & (0.00000) &  &  & (0.00000) \\ 
  & & & & \\ 
 grams & $-$0.009 &  &  & $-$0.004 \\ 
  & (0.010) &  &  & (0.010) \\ 
  & & & & \\ 
 CPU & 0.0003 & 0.00000 & 0.0002 & 0.001 \\ 
  & (0.001) & (0.001) & (0.0004) & (0.001) \\ 
  & & & & \\ 
 RAM & 0.0004 & 0.0003 &  & $-$0.003 \\ 
  & (0.002) & (0.001) &  & (0.003) \\ 
  & & & & \\ 
 GPU & $-$0.001 & $-$0.00004 & 0.002 & 0.015 \\ 
  & (0.009) & (0.004) & (0.002) & (0.015) \\ 
  & & & & \\ 
 Titles & $-$0.001 &  &  & 0.015 \\ 
  & (0.002) &  &  & (0.013) \\ 
  & & & & \\ 
 Exclusive &  & $-$0.003 & $-$0.002 & $-$0.034 \\ 
  &  & (0.004) & (0.003) & (0.026) \\ 
  & & & & \\ 
 Subscribe & $-$1.608 & $-$1.381 & $-$0.666 & 6.680 \\ 
  & (3.911) & (1.582) & (0.620) & (7.389) \\ 
  & & & & \\ 
 Price & $-$0.003 & $-$0.001 & $-$0.001 & $-$0.001 \\ 
  & (0.002) & (0.001) & (0.001) & (0.002) \\ 
  & & & & \\ 
\hline \\[-1.8ex] 
Observations & 22 & 22 & 22 & 22 \\ 
R$^{2}$ & 0.316 & 0.212 & 0.184 & 0.518 \\ 
Adjusted R$^{2}$ & $-$1.873 & $-$1.364 & $-$1.141 & $-$1.533 \\ 
F Statistic & 0.289 & 0.314 & 0.361 & 0.477 \\ 
\hline 
\hline \\[-1.8ex] 
\textit{Note:}  & \multicolumn{4}{r}{$^{*}$p$<$0.1; $^{**}$p$<$0.05; $^{***}$p$<$0.01} \\ 
\end{tabular} 
\end{table}

\clearpage

\begin{table}[!htbp] \centering 
  \caption[]{Two Stage Least Squares, IV estimation\footnotemark}
  \label{2SLS}
\begin{tabular}{@{\extracolsep{5pt}}lcccc} 
\\[-1.8ex]\hline 
\hline \\[-1.8ex] 
 & \multicolumn{4}{c}{\textit{Dependent variable:}} \\ 
\cline{2-5} 
\\[-1.8ex] & \multicolumn{4}{c}{log} \\ 
\\[-1.8ex] & (1) & (2) & (3) & (4)\\ 
\hline \\[-1.8ex] 
 Price & $-$0.002$^{**}$ & $-$0.001 & 0.002 & $-$0.001 \\ 
  & (0.001) & (0.001) & (0.002) & (0.001) \\ 
  & & & & \\ 
 CPU & 0.002$^{***}$ & 0.001$^{**}$ & 0.001 & 0.002$^{***}$ \\ 
  & (0.001) & (0.001) & (0.001) & (0.001) \\ 
  & & & & \\ 
 RAM & $-$0.001$^{***}$ & $-$0.001$^{***}$ & $-$0.001$^{***}$ & $-$0.001$^{***}$ \\ 
  & (0.0001) & (0.0001) & (0.0001) & (0.0001) \\ 
  & & & & \\ 
 GPU & 0.007$^{***}$ & 0.007$^{***}$ & 0.012$^{***}$ & 0.007$^{***}$ \\ 
  & (0.001) & (0.001) & (0.001) & (0.001) \\ 
  & & & & \\ 
 Titles &  & $-$0.002 & 0.008$^{***}$ & $-$0.001 \\ 
  &  & (0.001) & (0.002) & (0.001) \\ 
  & & & & \\ 
 Exclusive &  &  & $-$0.028$^{***}$ &  \\ 
  &  &  & (0.006) &  \\ 
  & & & & \\ 
 Storage &  & 0.001$^{*}$ & 0.002$^{**}$ &  \\ 
  &  & (0.001) & (0.001) &  \\ 
  & & & & \\ 
 Core &  &  & $-$0.039 &  \\ 
  &  &  & (0.060) &  \\ 
  & & & & \\ 
 Subscribe & 1.268$^{***}$ & 1.179$^{***}$ & 1.344$^{***}$ & 1.360$^{***}$ \\ 
  & (0.140) & (0.250) & (0.229) & (0.209) \\ 
  & & & & \\ 
 Constant & $-$1.366$^{***}$ & $-$1.359$^{***}$ & $-$1.263$^{***}$ & $-$1.395$^{***}$ \\ 
  & (0.354) & (0.299) & (0.186) & (0.312) \\ 
  & & & & \\ 
\hline \\[-1.8ex] 
Observations & 22 & 22 & 22 & 22 \\ 
R$^{2}$ & 0.715 & 0.744 & 0.854 & 0.719 \\ 
Adjusted R$^{2}$ & 0.626 & 0.615 & 0.745 & 0.607 \\ 
Residual Std. Error & 0.445 & 0.452 & 0.368 & 0.457 \\ 
\hline 
\hline \\[-1.8ex] 
\textit{Note:}  & \multicolumn{4}{r}{$^{*}$p$<$0.1; $^{**}$p$<$0.05; $^{***}$p$<$0.01} \\ 
\end{tabular} 
\end{table}
\footnotetext{The \texttt{robust.se} function was used from the ivpack R package to obtain heteroskedastic robust standard errors for instrumental variable analysis. i.e. These are the Huber-White standard errors for instrumental variable analysis as described in \cite{white1982instrumental}. \citep{ivpack}}

\newpage

\begin{table}[!htbp] \centering 
  \caption{Conditional $F$ test for (1) from Table \ref{2SLS}} 
  \label{F test (1)} 
\begin{tabular}{@{\extracolsep{5pt}}lccccccc} 
\\[-1.8ex]\hline 
\hline \\[-1.8ex] 
Statistic & \multicolumn{1}{c}{N} & \multicolumn{1}{c}{Mean} & \multicolumn{1}{c}{St. Dev.} & \multicolumn{1}{c}{Min} & \multicolumn{1}{c}{Pctl(25)} & \multicolumn{1}{c}{Pctl(75)} & \multicolumn{1}{c}{Max} \\ 
\hline \\[-1.8ex] 
Res.Df & 2 & 16.000 & 1.414 & 15 & 15.5 & 16.5 & 17 \\ 
Df & 1 & $-$2.000 &  & $-$2.000 & $-$2.000 & $-$2.000 & $-$2.000 \\ 
F & 1 & 8.712 &  & 8.712 & 8.712 & 8.712 & 8.712 \\ 
Pr(\textgreater F) & 1 & 0.003 &  & 0.003 & 0.003 & 0.003 & 0.003 \\ 
\hline \\[-1.8ex] 
\end{tabular} 
\end{table}

\begin{table}[!htbp] \centering 
  \caption{Conditional $F$ test for (2) from Table \ref{2SLS}} 
  \label{F test (2)} 
\begin{tabular}{@{\extracolsep{5pt}}lccccccc} 
\\[-1.8ex]\hline 
\hline \\[-1.8ex] 
Statistic & \multicolumn{1}{c}{N} & \multicolumn{1}{c}{Mean} & \multicolumn{1}{c}{St. Dev.} & \multicolumn{1}{c}{Min} & \multicolumn{1}{c}{Pctl(25)} & \multicolumn{1}{c}{Pctl(75)} & \multicolumn{1}{c}{Max} \\ 
\hline \\[-1.8ex] 
Res.Df & 2 & 14.000 & 1.414 & 13 & 13.5 & 14.5 & 15 \\ 
Df & 1 & $-$2.000 &  & $-$2.000 & $-$2.000 & $-$2.000 & $-$2.000 \\ 
F & 1 & 8.083 &  & 8.083 & 8.083 & 8.083 & 8.083 \\ 
Pr(\textgreater F) & 1 & 0.005 &  & 0.005 & 0.005 & 0.005 & 0.005 \\ 
\hline \\[-1.8ex] 
\end{tabular} 
\end{table}

\begin{table}[!htbp] \centering 
  \caption{Conditional $F$ test for (3) from Table \ref{2SLS}} 
  \label{F test (3)} 
\begin{tabular}{@{\extracolsep{5pt}}lccccccc} 
\\[-1.8ex]\hline 
\hline \\[-1.8ex] 
Statistic & \multicolumn{1}{c}{N} & \multicolumn{1}{c}{Mean} & \multicolumn{1}{c}{St. Dev.} & \multicolumn{1}{c}{Min} & \multicolumn{1}{c}{Pctl(25)} & \multicolumn{1}{c}{Pctl(75)} & \multicolumn{1}{c}{Max} \\ 
\hline \\[-1.8ex] 
Res.Df & 2 & 12.000 & 1.414 & 11 & 11.5 & 12.5 & 13 \\ 
Df & 1 & $-$2.000 &  & $-$2.000 & $-$2.000 & $-$2.000 & $-$2.000 \\ 
F & 1 & 2.529 &  & 2.529 & 2.529 & 2.529 & 2.529 \\ 
Pr(\textgreater F) & 1 & 0.125 &  & 0.125 & 0.125 & 0.125 & 0.125 \\ 
\hline \\[-1.8ex] 
\end{tabular} 
\end{table}

\begin{table}[!htbp] \centering 
  \caption{Conditional $F$ test for (4) from Table \ref{2SLS}} 
  \label{F test (4)} 
\begin{tabular}{@{\extracolsep{5pt}}lccccccc} 
\\[-1.8ex]\hline 
\hline \\[-1.8ex] 
Statistic & \multicolumn{1}{c}{N} & \multicolumn{1}{c}{Mean} & \multicolumn{1}{c}{St. Dev.} & \multicolumn{1}{c}{Min} & \multicolumn{1}{c}{Pctl(25)} & \multicolumn{1}{c}{Pctl(75)} & \multicolumn{1}{c}{Max} \\ 
\hline \\[-1.8ex] 
Res.Df & 2 & 15.000 & 1.414 & 14 & 14.5 & 15.5 & 16 \\ 
Df & 1 & $-$2.000 &  & $-$2.000 & $-$2.000 & $-$2.000 & $-$2.000 \\ 
F & 1 & 9.222 &  & 9.222 & 9.222 & 9.222 & 9.222 \\ 
Pr(\textgreater F) & 1 & 0.003 &  & 0.003 & 0.003 & 0.003 & 0.003 \\ 
\hline \\[-1.8ex] 
\end{tabular} 
\end{table} 

\begin{table}[!htbp] \centering 
  \caption{Residual Regression} 
  \label{residual regression} 
\begin{tabular}{@{\extracolsep{5pt}}lc} 
\\[-1.8ex]\hline 
\hline \\[-1.8ex] 
 & \multicolumn{1}{c}{\textit{Dependent variable:}} \\ 
\cline{2-2} 
\\[-1.8ex] & res.IV1 \\ 
\hline \\[-1.8ex] 
 CPU\_cost & 0.005 \\ 
  & (0.004) \\ 
  & \\ 
 RAM\_cost & $-$0.006 \\ 
  & (0.009) \\ 
  & \\ 
 CPU & 0.001 \\ 
  & (0.001) \\ 
  & \\ 
 RAM & $-$0.00002 \\ 
  & (0.0002) \\ 
  & \\ 
 GPU & $-$0.001 \\ 
  & (0.002) \\ 
  & \\ 
 Subscribe & 0.163 \\ 
  & (0.444) \\ 
  & \\ 
 Constant & $-$0.296 \\ 
  & (0.475) \\ 
  & \\ 
\hline \\[-1.8ex] 
Observations & 22 \\ 
R$^{2}$ & 0.112 \\ 
Adjusted R$^{2}$ & $-$0.243 \\ 
Residual Std. Error & 0.430 (df = 15) \\ 
F Statistic & 0.315 (df = 6; 15) \\ 
\hline 
\hline \\[-1.8ex] 
\textit{Note:}  & \multicolumn{1}{r}{$^{*}$p$<$0.1; $^{**}$p$<$0.05; $^{***}$p$<$0.01} \\ 
\end{tabular} 
\end{table}

\renewcommand\thesubsection{\Alph{subsection}}

\clearpage
\appendix
\section*{Appendix}
\subsection{Logit model for probabilistic choice}\label{Appendix A}
First we define $\tilde{\varepsilon} \equiv \varepsilon_{ij't}-\varepsilon_{ijt}$. Where $\tilde{\varepsilon}$ will follow a logistic CDF by assumption 1. in Section \ref{Estimation}. Given the indirect random utility from (\ref{indirect random utility}), 

\begin{equation}
\begin{split}
\Pr(\text{Consumer chooses }j) & = \Pr(\delta_{jt} + \varepsilon_{ijt} > \delta_{j't} + \varepsilon_{ij't}) \\
& = \Pr(\varepsilon_{ij't}-\varepsilon_{ijt} < \delta_{jt} - \delta_{j't}) \\
& = \Pr(\tilde{\varepsilon} < \delta_{jt} - \delta_{j't}) \\
& = \frac{\exp \delta_{jt}}{\exp \delta_{jt}+\exp \delta_{j't}}
\end{split} 
\end{equation}

\subsection{Dataset}
Dataset and code can be accessed via. \\
\url{https://github.com/ctymarco/gamepass/tree/10c6983afdf08ba8123cecf162f34fd218b76e61}

\clearpage

\bibliographystyle{apacite}
\bibliography{ref}

\end{document}